\documentclass[%
 twocolumn,amsmath,amssymb
]{revtex4-1}

\usepackage{graphicx}
\usepackage{dcolumn}
\usepackage{bm}

\usepackage[utf8]{inputenc}
\usepackage{etoolbox}

\usepackage{algorithm}
\usepackage{algorithmic}

\usepackage{hyperref}
\usepackage[capitalise]{cleveref}

\usepackage{multirow}

\usepackage{xcolor}
\renewcommand{\H}{\textbf{H}}
\newcommand{\C}{\textbf{C}}

\newcommand{\F}{\textbf{F}}

\renewcommand{\P}{\textbf{P}}

\newcommand{\boldrho}{\text{\boldmath$\rho$}}
\newcommand{\bg}{\text{\boldmath$\gamma$}}
\newcommand{\bG}{\text{\boldmath$\Gamma$}}
\newcommand{\bL}{\text{\boldmath$\Lambda$}}
\newcommand{\Tr}{\text{Tr}}
\newcommand{\U}{\textbf{U}}
\renewcommand{\k}{\textbf{k}}

\makeatletter
\def\@email#1#2{%
 \endgroup
 \patchcmd{\titleblock@produce}
  {\frontmatter@RRAPformat}
  {\frontmatter@RRAPformat{\produce@RRAP{*#1\href{mailto:#2}{#2}}}\frontmatter@RRAPformat}
  {}{}
}%
\makeatother
\begin{document}

\title{Gaussian-Based Periodic Grand Canonical Density Functional Theory with Implicit Solvation for Computational Electrochemistry}

\author{Anton Z. Ni}
\author{Adam Rettig}
\author{Joonho Lee$^*$}
 \email{joonholee@g.harvard.edu}
\affiliation{ 
Department of Chemistry and Chemical Biology, Harvard University, Cambridge, MA
}

\date{\today}

\begin{abstract}
We present a numerical method for grand canonical density functional theory (DFT) tailored to solid-state systems, employing Gaussian-type orbitals as the primary basis. Our approach directly minimizes the grand canonical free energy using the density matrix as the sole variational parameter, while self-consistently updating the electron number between self-consistent field iterations. To enable realistic electrochemical modeling, we integrate this approach with implicit solvation models. Our solvation scheme introduces less than 50\% overhead relative to gas-phase calculations. Compared to existing plane wave-based implementations, our method shows improved robustness in grand canonical simulations. We validate the approach by modeling corrosion at silver surfaces, finding excellent agreement with previous studies. Our method is implemented in the quantum chemistry software Q--Chem. This work lays the groundwork for future wavefunction-based simulations beyond DFT under electrochemical operando conditions.
\end{abstract}

\maketitle

\section{Introduction}
Electrochemical reactions at solid-liquid interfaces are the cornerstone of many critical processes for renewable energy, such as carbon dioxide reduction and water splitting.\cite{walter2011,seh2017} 
As a result, there has been a considerable interest in designing more efficient electrochemical cells and catalysts to perform these reactions.\cite{gur2018} 
In parallel with experimental progress, there is a growing demand for accurate computational modeling of the chemistry and physics occurring at solid-liquid interfaces.\cite{norskov2011,chen2019,hammesschiffer2021,ringe2021} Because interfacial electrochemical reactions inherently involve electron transfer---often accompanied by bond formation and breaking---accurately modeling these processes requires a quantum mechanical description of the electrons at the interface.

One challenge of quantum mechanically modeling solid-liquid interfaces is that an atomistic model is computationally infeasible even with a relatively economical method such as density functional theory (DFT).\cite{bruix2019} 
Fortunately, chemical reactivity in these systems is often localized to the environment near the interface. 
Therefore, only the portion of the electrode close to the surface needs to be explicitly modeled. 
To maintain computational tractability, the solvent and electrolyte environment can be coarse-grained down to a polarizable continuum with an implicit solvation model.\cite{ringe2021} 
These methods in general work by defining a solvent-accessible cavity that influences the electrostatic potential in the solute environment. 
Ionic screening is also included to account for the influence of electrolyte ions in solution. 
These methods have been implemented in a wide range of periodic quantum chemistry codes based on plane waves.\cite{mathew2019,fisicaro2016,letchworth-weaver2012,dziedzic2020}

Moreover, electrochemical reactions are often performed under a constant electrochemical potential. Under these conditions, one aims to minimize the grand canonical free energy variationally, as opposed to the minimization of the electronic energy. This is challenging as most electronic structure methods available to date are formulated for canonical ensembles and assume constant electron number. 
In some past works, a ``constant potential" calculation was performed by running multiple fixed-electron-number calculations and interpolating to obtain the proper chemical potential.\cite{bonnet2012,gauthier2019} 
These methods require additional computational or manual effort, so a technique that avoids these extra complications is highly desirable. 
For this reason, previously, direct minimization methods with an auxiliary Hamiltonian and related methods have been developed in plane wave density functional theory codes.\cite{marzari1997,freysoldt2009,sundararaman2017,melander2018nov} 
We also note a few exceptions for many-body methods such as perturbation theory,\cite{hirata2020jul,wei2022jul} coupled-cluster methods,\cite{baeurle2002aug,hummel2018dec,white2018nov,white2020jun,harsha2019nov} quantum Monte Carlo,\cite{shen2020nov,malone2015jul,lee2021feb} etc., where grand canonical electronic structure methods were developed focusing on studying warm dense matter in the gas phase.

Traditionally, periodic electronic structure calculations are performed with plane waves as the basis of choice.\cite{kresse1996,qe,jdftx} This is because plane waves are inherently periodic and form an orthogonal basis set, making them free from basis set superposition error (BSSE). 
On the other hand, Gaussian-type orbitals (GTOs), the natural choice for molecular applications, are non-orthogonal and local. They offer a more compact representation than plane waves for performing many-body calculations.
Therefore, despite their deficiencies, such as BSSE and basis set linear dependencies,\cite{klahn1977,vilelaoliveira2019} GTOs offer unique strengths, especially beyond pure density functional calculations.\cite{cryscor,pisani2008,rybkin2016may,robinson2025}
Recently, work has been done to develop GTO basis sets that extrapolate well to the basis set limit, making GTOs a much more viable choice as the computational basis in solids.\cite{daga2020,peintinger2013,ye2022}

This paper is organized as follows. We first present our implementation of periodic grand canonical density functional theory (GCDFT) in a GTO-based code that adaptively updates the electron number between self-consistent field (SCF) iterations. Our method performs a variational minimization of the grand potential over density matrices, as initially proposed by Mermin in the celebrated thermal Hartree-Fock theory paper.\cite{mermin1963} Such a direct implementation is impractical in plane wave frameworks due to the significant storage overhead associated with density matrices. Furthermore, our method does not require simultaneous optimization of both molecular orbital coefficients and an auxiliary Hamiltonian used in plane wave implementations, which, as we demonstrate later, shows faster convergence than preexisting plane wave methods.\cite{marzari1997,freysoldt2009,sundararaman2017} 

We also present the implementation of two implicit solvation models, which incorporate linear dielectric and ionic response into the density functional. Both models have demonstrated accuracy on a wide range of solutes and are seamlessly integrated into our GCDFT implementation.\cite{gunceler2013,candle}
We note that while CP2K has an implementation of grand canonical density functional theory with the self-consistent continuum solvation (SCCS) model,\cite{andreussi2012} they employ a Pulay mixing scheme rather than a more robust variational minimization method.\cite{cp2k,chai2024sep} The  CP2K implementation uses a planar counter charge formalism for ionic screening rather than a more realistic implicit electrolyte.
To the best of our knowledge, no existing GTO-based implementation supports variational minimization of the grand canonical free energy in the presence of solvation under periodic boundary conditions. 
Our resulting GCDFT method enables realistic simulations of electrochemical systems. 
We present an application of our method to a detailed study of the corrosion of a silver electrode surface.\cite{kang2024feb} 
This work demonstrates the viability, robustness, and potential of using GTOs as a computational basis for practical problems in electrochemistry, which is currently dominated by plane wave-based codes.\cite{goodpaster2016apr,garza2018feb}

\section{Theoretical Framework and Implementation}
\subsection{Grand Canonical Density Functional Theory}
We outline a DFT calculation at a constant (electro)chemical potential or in a grand canonical ensemble. Here, because the electron number is no longer held constant, the grand canonical free energy is variationally minimized as opposed to the total electronic energy.
The grand canonical free energy is defined as
\begin{equation}
    \Omega = E_\text{DFT} - \mu N - \frac{1}{\beta}S_\text{el},
\label{eq:grandpot}
\end{equation}
where $E_\text{DFT}$ is the DFT electronic energy, $\mu N$ is the electron number scaled by the fixed chemical potential $\mu$, $\beta$ is the inverse temperature, and $S_\text{el}$ is the electronic entropic energy.

\noindent We discretize the underlying operators in \cref{eq:grandpot} with orthonormalized GTOs,
\begin{widetext}
    \begin{equation}
    \frac{\Omega}{N_k} = \frac{1}{N_k}\sum_{\mathbf k}
    \left(\Tr({\P^{\mathbf k}}({\textbf{h}^{\mathbf k}}+\frac{1}{2}{\textbf{J}^{\mathbf k}})) + E_\text{xc}[\rho(\mathbf r),\cdots] - \mu \Tr({\P}^{\mathbf k}) + \frac{1}{\beta} S_{\text{el}}^\mathbf k\right),
    \label{eq:discOmega}
\end{equation}
\end{widetext}
where ${\mathbf k}$ is a $\mathbf k$-point in the Brillouin zone, $N_k$ is the number of $\mathbf k$-points, ${\P}^{\mathbf k}$ is the one-body reduced density matrix (1RDM) at $\mathbf k$, ${\textbf{h}}^{\mathbf k}$ and ${\textbf{J}}^{\mathbf k}$ are the one-body Hamiltonian matrix and the Coulomb matrix at $\mathbf k$, respectively, $E_\text{xc}$ is the exchange-correlation energy, and $S_{\text{el}}^{\mathbf k} = \Tr({\P}^{\mathbf k}\log({\P}^{\mathbf k}) + (\textbf{I}-{\P}^{\mathbf k})\log(\textbf{I}-{\P}^{\mathbf k}))$.
In practice, we use the transformation between non-orthogonal atomic orbitals and orthogonalized atomic orbitals to correctly handle the metric when evaluating \cref{eq:discOmega} with ${\textbf{h}}^{\mathbf k}$ and ${\textbf{F}}^{\mathbf k}$ evaluated in the atomic orbital basis.

If we constrain the 1RDM to be Hermitian with eigenvalues between 0 and 1, the DFT thermal state is obtained by minimizing $\Omega$ over all such 1RDMs. 
To impose these constraints during the optimization, we introduce a matrix ${\H}$, which parametrizes 
\begin{equation}
{\P}^{\mathbf k} = \frac{1}{\exp(\bG^{\mathbf k}) + 1} = \U^{\mathbf k}\frac{1}{\exp(\bg^{\mathbf k})+1}\U^{\mathbf k H} = \U^{\mathbf k}\boldrho^{\mathbf k}\U^{\mathbf k H},
\label{eq:pfromh}
\end{equation}
where $\bG^{\mathbf k} = \beta\left(\frac{1}{2}\left( {\H}^{\mathbf k} + {\H}^{\mathbf k H} \right) - \mu \textbf{I} \right) = \U^{\mathbf k}\bg^{\mathbf k}\U^{\mathbf k H}$. We can then minimize the free energy expression by performing an unconstrained minimization with respect to $\{{\H}^{\mathbf k}\}$. At convergence, we have ${\H}^{\mathbf k} = {\F}^{\mathbf k}$. We note that this method differs from the popular auxiliary Hamiltonian approaches in that we do not optimize the coefficient matrix as a separate variational parameter.\cite{marzari1997,freysoldt2009,sundararaman2017} Instead, our molecular orbital coefficient matrix, $\{\mathbf C^{\mathbf k}\}$, is determined by diagonalizing $\bG^{\mathbf k}$. Additionally, due to the compactness of GTOs, we can work with the full density matrix rather than a subspace of the given basis set as is done in plane wave-based codes. This is because in plane wave-based codes, the number of basis functions is typically so large that it becomes infeasible to store the full density matrix. Our method incurs minimal storage overhead, as \{\textbf{H}\} has the same dimension as the density matrix. While a recent effort was made to perform direct minimization in the presence of fluctuating orbital population, the method is applicable only to systems with integer electrons.\cite{li2025may} 

We perform the minimization of the free energy using the nonlinear conjugate gradient method. In the method, we use the residual $\frac{1}{2}({\F}^{\mathbf k} - {\H}^{\mathbf k})$ as the preconditioned gradient.\cite{sundararaman2017} We note that this method converges to the true minimum since ${\F}^{\mathbf k} - {\H}^{\mathbf k} = 0$ at convergence. The gradient of the objective function $\Omega$ with respect to $\bG^{\mathbf k}$ can be computed efficiently, and the gradient with respect to ${\H}^{\mathbf k}$ is obtained by symmetrizing the gradients with respect to $\bG^{\mathbf k}$. 
The detailed forms of the gradients are included below.  
All expressions are written using the Einstein summation convention. We define $G_{li}^\k := \frac{1-\delta_{li}}{\gamma_{i}^\k - \gamma_{l}^\k}$ where $\gamma_{i}^\k$ is the $i$-th eigenvalue of $\mathbf \bG^{\mathbf k}$.
\begin{widetext}
\begin{gather}
    \begin{split}
        \frac{\partial \Omega}{\partial \text{Re}(\Gamma^\k_{pq})} =\quad &{F}^\k_{mn}\left( U_{pl}^{\k *}U^\k_{qi}U_{nl}^\k G_{li}^\k \rho_i^\k U_{mi}^{\k *} + U_{ni}^\k \rho_i^\k U_{pl}^\k U_{qi}^{\k *} U_{ml}^{\k *} G_{li}^\k - U_{ni}^\k U_{mi}^{\k *}\rho_i^\k (1-\rho_i^\k)U_{pi}^{\k *} U_{qi}^{\k} \right) \\
        &+ \mu (U_{pi}^{\k *} \rho_i^\k (1-\rho_i^\k)U_{qi}^{\k}) + \frac{1}{\beta}\gamma_i^\k \rho_i^\k(1-\rho_i^\k)U_{pi}^{\k *} U_{qi}^{\k}
    \end{split}\\
    \begin{split}
        \frac{\partial \Omega}{\partial \text{Im}(\Gamma^\k_{pq})} = \quad  &i{F}^\k_{mn}\left( U_{pl}^{\k *}U^\k_{qi}U_{nl}^\k G_{li}^\k \rho_i^\k U_{mi}^{\k *} - U_{ni}^\k \rho_i^\k U_{pl}^\k U_{qi}^{\k *} U_{ml}^{\k *} G_{li}^\k - U_{ni}^\k U_{mi}^{\k *}\rho_i^\k (1-\rho_i^\k)U_{pi}^{\k *} U_{qi}^{\k} \right) \\
        &+ i\mu (U_{pi}^{\k *} \rho_i^\k (1-\rho_i^\k)U_{qi}^{\k}) + i\frac{1}{\beta}\gamma_i^\k \rho_i^\k(1-\rho_i^\k)U_{pi}^{\k *} U_{qi}^{\k}
    \end{split}
\end{gather}
\end{widetext}

\noindent To get the gradient with respect to ${\H}^\k$ from the gradient with respect to $\bG^\k$, we can use the following: 
\begin{equation}
    \frac{\partial \Omega}{\partial \text{Re}({\H}^\k)} = \frac{\beta}{2}\left(\frac{\partial\Omega}{\partial \text{Re}\left(\bG^\k\right)} + \left( \frac{\partial\Omega}{\partial \text{Re}\left(\bG^\k\right)} \right)^T\right) 
\end{equation}
\begin{equation}
    \frac{\partial \Omega}{\partial \text{Im}({\H}^\k)} = \frac{\beta}{2}\left(\frac{\partial\Omega}{\partial \text{Im}\left(\bG^\k\right)} - \left( \frac{\partial\Omega}{\partial \text{Im}\left(\bG^\k\right)} \right)^T\right)
\end{equation}
\noindent With this method, we can converge systems at a wide range of chemical potentials. Our method has the same computational cost per iteration as standard electronic-structure convergence techniques, since the dominant cost remains the construction of the Fock matrix \textbf{F}, which is independent of the convergence scheme.

Note that this method can also be used to perform fixed-electron calculations as in canonical density functional theory by adding the gradient of the chemical potential $\mu$ with respect to the variational parameters and using the canonical free energy $E = E_\text{DFT} - TS_\text{el}$ as the objective function for optimization.\cite{freysoldt2009} This algorithm was also used to obtain the potential of zero charge (PZC) of metal slabs later in this work.
The relevant gradients are as follows:
\begin{align}
    \frac{\partial \mu}{\partial\text{Re}({H}^{\mathbf k}_{pq})} &= \frac{(U_{pi}^{\mathbf k})^*\rho_i^{\mathbf k}(1-\rho_i^{\mathbf k})U_{qi}^{\mathbf k}}{\sum_j \rho_j^{\mathbf k}(1-\rho_j^{\mathbf k})}\\
    \frac{\partial \mu}{\partial\text{Im}({H}_{pq}^{\mathbf k})} &= i\frac{(U_{pi}^{\mathbf k})^*\rho_i^{\mathbf k}(1-\rho_i^{\mathbf k})U_{qi}^{\mathbf k}}{\sum_j \rho_j^{\mathbf k}(1-\rho_j^{\mathbf k})}.
\end{align}

\subsection{Implicit Solvation Framework}

Meaningful constant potential calculations in periodic systems must resolve one problem inherent to periodic boundary conditions. That is, the potential energy of charged unit cells in an infinite lattice diverges. In nearly all periodic electronic structure codes, the default way to handle this is by zeroing out the $G = 0$ component of the electrostatic potential in reciprocal space, which is equivalent to introducing a uniform background charge that neutralizes the charge of the system.\cite{qe,kresse1996} It is known that the choice of neutralizing charge can affect qualitative predictions, so a physically informed choice should be used in calculations.\cite{bhandari2020sep} 
As electrochemical reactions are performed in the presence of an electrolyte solution, the most logical way to enforce the net charge neutrality is by introducing ionic screening provided by electrolytes in solution. 
It is well-known that introducing linear ionic response (i.e., Debye screening) is already sufficient to neutralize the unit cell exactly.\cite{letchworth-weaver2012} 

In this work, we have implemented a density-based polarizable continuum solvation model (PCM), which incorporates linear dielectric and ionic response. We demonstrate the implementation of two cavity functions. The first is a locally determined cavity based on the local linear PCM model (LPCM) in JDFTx and VASPsol.\cite{gunceler2013,mathew2019} The LPCM model is equivalent to computing the solvation potential via the linearized Poisson-Boltzmann equation.
Secondly, we have implemented the charge-asymmetric nonlocally-determined local-electric (CANDLE) cavity function, known for its prediction of accurate energies for charged solutes, especially anions, which are systematically undersolvated in most PCM models.\cite{candle}
We provide more implementation details of these solvation models below.

\subsubsection{Solvated DFT energy}
In general, the solvated DFT energy can be decomposed in the following form:\cite{vaspsol++}
\begin{equation}
    E_\text{DFT}[\rho,\phi]  = E_T + E_J[\rho] + E_{\text{xc}, \text{gas}}[\rho] + 
    E_\text{solv}[\rho,\phi] ,
\end{equation}
where the solvation energy is
\begin{equation}
E_\text{solv} [\rho,\phi]  = E_{J,\text{solv}}[\rho,\phi] + E_\text{diel}[\rho,\phi] + E_\text{ion}[\rho,\phi] + E_\text{cd}[\rho],
\label{eq:esolv}
\end{equation}
with $E_\text{gas}$ representing the energy in the absence of solvation, $E_{J,\text{solv}}$ being the Coulomb correction, $E_\text{diel}$ being the dielectric free energy, $E_\text{ion}$ being the ionic free energy, and $E_\text{cd}$ corresponding to nonelectrostatic cavitation and dispersion energies. 
There are two variational parameters: $\rho(\textbf{r})$ and $\phi(\textbf{r})$ which are the solute valence electron density and the overall electrostatic potential in the fluid environment, respectively. Here, $\rho$ represents the valence electronic density since the core is absent due to pseudopotentials. Later, we will use $\rho_\text{sol}$, $\rho_\text{diel}$, $\rho_\text{ion}$, $\rho_\text{tot}$, to represent the sum of nuclear and valence electronic density, bound solvent charge density, ionic charge density, and total charge density, respectively. $\phi_\text{sol}$ and $\phi\equiv \phi_\text{tot}$ are defined analogously for the electrostatic potential. We define $\phi_\text{solv} := \phi_\text{tot} - \phi_\text{sol}$. The form of each energy component of $E_\text{solv}$ under linear fluid response is given in this section.

\subsubsection{Cavity functions}
In continuum solvation frameworks, the solvent-accessible cavity function $s(\textbf{r})$ must be defined. This function is usually a twice-differentiable switching function that is zero in the solute and one in the continuum region. There are many possible formulations of the cavity functions.\cite{letchworth-weaver2012,andreussi2012,fisicaro2017,stein2019} The cavity function defines the dielectric cavity $s_\epsilon(\textbf{r})$ which can vary between solvation models. The dielectric cavity defines the effective dielectric constant over the grid in these polarizable continuum models such that the dielectric constant is 1 in the solute region and is $\epsilon_b$, the bulk dielectric constant, in the solvent region.
\begin{equation}
    \epsilon(\textbf{r}) = 1 + (\epsilon_b -1)s_\epsilon(\textbf{r})
\end{equation}

In our implementation, we followed the density-based cavity functions adopted by JDFTx and VASPsol.\cite{gunceler2013,mathew2019} These are known to be efficient and accurate cavities for most systems and are routinely used in computational electrocatalysis problems. In the local and linear polarizable continuum model (LPCM), the local density cavity function is parametrized by two tunable parameters, $\rho_\text{cut}$ and $\sigma$. We use $\rho_\text{cut} = 0.00037$ $a_0^{-3}$ and $\sigma = 0.6$.\cite{gunceler2013}
\begin{equation}
    s_{\epsilon,\text{LPCM}} = s_\text{LPCM}(\textbf{r}) = \frac{1}{2}\text{erfc}\left( \frac{\log(\rho(\textbf{r})/\rho_\text{cut})}{\sigma\sqrt{2}} \right)
\end{equation}

We have also implemented the charge-asymmetric nonlocally determined (CANDLE) shape function, which incorporates charge asymmetry into the shape function.\cite{candle} Nonlocality is incorporated into the CANDLE cavity through the convolution of the sphericalized density of a solvent molecule $w_\text{lq}$ and the valence electronic density $\rho(\textbf{r})$. This convolution defines an effective electronic density to build the cavity: $\overline{\rho}(\textbf{r}) = (w_\text{lq} * \rho)\;(\textbf{r})$. Unique to CANDLE is the presence of a charge asymmetric term, which moves the solvation cavity closer to the solute for anions.\cite{candle} 
To obtain the dielectric cavity, the cavity function is smeared via convolution with a radial delta function so that the transition from solute to bulk solvent is less steep. The cavity function is parametrized by a solvent-dependent quantity $p_\text{cav}$, the coefficient on the charge asymmetric term. For water, $p_\text{cav}$ is 36.5. The quantity $Z_\text{val}$ is the effective solute nuclear charge, and $\hat{e}_{\nabla\overline{\rho}}$ is the unit vector parallel to $\nabla\overline{\rho}$.
\begin{equation}
    \begin{split}
    s_\text{CANDLE}(\textbf{r}) =\; &\frac{1}{2}\text{erfc}\left( \log\frac{Z_\text{val}\overline{\rho}(\textbf{r})}{0.00142} \right. \\ 
     &- \left. \text{sign}(p_\text{cav})f_\text{sat}\left( |p_\text{cav}|\hat{e}_{\nabla\overline{\rho}} \cdot \nabla\overline{\phi}_\text{sol} \right) \right)
    \end{split}
\end{equation}
\begin{equation}
    f_\text{sat}(x) =
\begin{cases}
0, & x < 0, \\
3\tanh\left(x^2\right), & x \ge 0
\end{cases}
\end{equation}
\begin{equation}
    s_{\epsilon,\text{CANDLE}}(\textbf{r}) = \frac{\delta(|\textbf{r}| - R_{\epsilon})}{4\pi R_{\epsilon}^2} * s_\text{CANDLE}(\textbf{r})
\end{equation}

For the computation of all cavity functions, a Gaussian core charge density is added to the valence electronic density at each solute atomic site to prevent solvent leakage into nuclear cores.

\subsubsection{Coulomb energy correction}
The Coulomb energy correction accounts for the difference in electronic Coulombic energy due to the difference between the electrostatic potential in solution versus in the gas phase. The Coulomb energy correction is shown below, 
\begin{equation}
E_{J,\text{solv}}[\rho,\phi] = \frac{q_\text{sol}}{V}\int d^3\textbf{r}\; \phi_\text{solv} + \frac{1}{8\pi}\int d^3\textbf{r}\; \phi_\text{solv}\nabla^2\phi_\text{solv},
\end{equation}
where $q_\text{sol}$ is the charge of the solute, and $V$ is the volume of the unit cell.

\subsubsection{Dielectric free energy}
\noindent Within linear dielectric response, the dielectric free energy can be expressed as 
\begin{equation}
    E_\text{diel}[\rho,\phi] = \int d^3\textbf{r}\; \frac{1}{2}\rho_\text{diel}(\textbf{r})\phi(\textbf{r})
\end{equation}
where $\rho_\text{diel}(\textbf{r}) = \frac{\epsilon_b - 1}{4\pi}\nabla(s_\epsilon(\textbf{r})\cdot\nabla\phi(\textbf{r}))$. If we plug this into the integral and integrate by parts, we obtain 
\begin{equation} E_\text{diel}[\rho,\phi] = -\int d^3\textbf{r}\; \frac{(\epsilon_b-1)s_\epsilon(\textbf{r})}{8\pi}|\nabla\phi(\textbf{r})|^2.\end{equation}

\subsubsection{Ionic free energy}
\noindent Within linear ionic response, the ionic free energy is given by \begin{equation} E_\text{ion}[\rho,\phi] = \int d^3\textbf{r}\; \frac{1}{2}\rho_\text{ion}(\textbf{r})\phi(\textbf{r}).\end{equation} 
The ionic charge density is given by $\rho_\text{ion} \equiv -\frac{\epsilon_b\kappa^2}{4\pi}s_\epsilon(\textbf{r})\phi(\mathbf{r})$ where $\kappa$ is the inverse Debye screening length. The Debye screening length can be computed as a function of the bulk concentration of each ionic species $c_i$ via \begin{equation}
    \epsilon_b\kappa^2 = 4\pi\beta\sum_i z_i^2c_i,
\end{equation}
where $z_i$ is the charge of the $i$-th ionic species.

\subsubsection{Nonelectrostatic energy}

The final term in the free energy is associated with cavitation and dispersive forces. In most solvation models, the cavitation energy is approximated by scaling the surface area of the solvent accessible cavity with $\tau$, a parameter that resembles surface tension.\cite{scherlis2006,andreussi2012,ringe2016} This is the form we use for our LPCM implementation, where we have $\tau = 5.4\times 10^{-6}\; E_h\cdot a_0^{-2}$: \begin{equation} E_\text{cav, LPCM}[\rho] = \tau\int d^3\textbf{r}\; |\nabla s(\textbf{r})|. \end{equation}
In the CANDLE solvation model, a parameter-free nonlocally determined cavitation energy and solvent D2-type interactions comprise $E_\text{cd}$. \cite{sundararaman2014oct,grimme-d2} 

\subsubsection{Implementation details}
The equation for the solvation potential can be computed by taking the variation of free energy with respect to $\phi$. Upon differentiation, we obtain the linearized Poisson-Boltzmann equation, which can be solved efficiently in reciprocal space using the preconditioned conjugate gradient algorithm.\cite{letchworth-weaver2012,mathew2019} 
\begin{equation} -\nabla(\epsilon(\textbf{r})\cdot\nabla\phi(\textbf{r}))+ \epsilon_\text{b}\kappa^2s_\epsilon(\textbf{r})\phi(\textbf{r}) =4\pi\rho_\text{sol}(\textbf{r}). \label{eq:pb}\end{equation}

\noindent By taking the variation of free energy with respect to $\rho$, we obtain the correction to the Kohn-Sham potential that can be used to determine the solvation energy self-consistently: \begin{equation} V_\text{solv} = \phi_\text{solv} - \left(\frac{\epsilon_\text{b}-1}{8\pi}\right)\frac{\delta s_\epsilon}{\delta \rho}|\nabla\phi|^2 - \beta c_\text{ion}z_\text{ion}^2\frac{\delta s_\epsilon}{\delta \rho}\phi^2 + \frac{\delta E_\text{cd}}{\delta\rho}. \label{eq:vsolv}\end{equation}

\noindent At each SCF iteration, we solve for the updated total electrostatic potential $\phi$ based on the new density. This procedure does not change whether we are performing a constant electron number or a constant potential calculation. 
Our implementation of the implicit solvent method requires little additional memory, since only the electronic density, its derived quantities, and the electrostatic potential need to be stored on a uniform grid. The density and its gradients are already evaluated on this grid for the computation of exchange-correlation contributions.
The general workflow of our method is detailed in Algorithm~\ref{alg:scf-loop}. For completeness, we provide the full derivation of $\delta E/\delta\rho$ in the Supporting Information.
\begin{algorithm}[H]
    \caption{Workflow for each SCF iteration of GCDFT} \label{alg:scf-loop}
    \begin{algorithmic}
    \FOR{each electronic SCF iteration}
        \STATE Build $\textbf{P}$ given $\textbf{H}$ (\cref{eq:pfromh}).
        \STATE Obtain $\rho(\textbf{r})$ from $\textbf{P}$.
        \STATE Compute $\phi_\text{sol}$ by solving $-\frac{1}{4\pi}\nabla^2\phi_\text{sol} = \rho(\textbf{r}) - \rho_\text{nuc}(\textbf{r})$.
        \STATE Compute cavity functions $s(\textbf{r})$ and $s_\epsilon(\textbf{r})$.
        \STATE Solve \cref{eq:pb} for $\phi$.
        \STATE Solve for $V_\text{solv}$ using \cref{eq:vsolv} using $\phi$.
        \STATE Build the Fock matrix $\textbf{F}_\text{solv}$ using $V_\text{solv}$.
        \STATE $\textbf{F} \leftarrow \textbf{F} + \textbf{F}_\text{solv}$.
        \STATE Obtain gradients $\frac{\partial\Omega}{\partial \textbf{H}}$ and residual $\textbf{F}-\textbf{H}$.
        \STATE Perform a line search along the search direction defined by the nonlinear conjugate gradient method.
        \STATE Update $\textbf{H}$.
    \ENDFOR
    \end{algorithmic}
\end{algorithm}

\section{Computational Details}
Both the grand canonical DFT method and implicit solvation models are implemented in a development version of Q--Chem with Periodic Boundary Conditions (QCPBC).\cite{qchem,lee2021oct,lee2022dec,rettig2023sep} Our implementation supports shared-memory parallel execution across multiple CPU cores. The code uses the Goedecker-Teter-Hutter (GTH) pseudopotentials optimized for the Perdew-Burke-Ernzerhof (PBE) functional and the corresponding optimized basis sets for the GTH pseudopotentials.\cite{pbe,gth,vandevondele2007} Unless otherwise noted, the TZVP-MOLOPT-PBE-GTH basis set developed by CP2K was used for all calculations.\cite{cp2k} The Gaussian and Plane Waves method (GPW) was used for the evaluation of two-body integrals.\cite{gpw} Unless otherwise specified, all metallic slab calculations were performed with a five-layer slab with one atom per layer with 15 \AA\ of vacuum. A $12\times 12\times 1$ $\Gamma$-centered $\mathbf k$-point mesh was used for such calculations. Fermi smearing with a smearing temperature of 0.001 a.u. was employed for all metallic calculations. All calculations were performed on an AMD EPYC 9654 Processor where 16 cores and 64 GB of memory were allocated.

\section{Results and Discussion}

\subsection{Verification of Implemented Solvation Models}

\begin{table}
    \centering
    \begin{tabular}{lccc}
    \hline\hline
    \noalign{\vskip 0.3em}
    \multirow{2}{*}{Model} & \multicolumn{3}{c}{MAE (kcal/mol)} \\
     & Neutral & Cations & Anions \\[0.3em]
    \hline
    \noalign{\vskip 0.3em}
    JDFTx LPCM & 1.27 & 2.10 & 15.09 \\[0.3em]
    QCPBC LPCM         & 1.15 & 2.23 & 13.91 \\[0.3em]
    JDFTx CANDLE       & 1.27 & 2.62 &  3.46 \\[0.3em]
    QCPBC CANDLE       & 1.17 & 2.70 &  3.57 \\[0.3em]
    \hline\hline
    \end{tabular}
    \caption{Comparison of the accuracy of LPCM and CANDLE between JDFTx (plane waves basis set with ultrasoft pseudopotentials)\cite{jdftx} and QCPBC (our implementation). Errors for JDFTx LPCM\cite{sundararaman2017feb} and JDFTx CANDLE\cite{candle} are obtained from corresponding references.}
    \label{tab:solv-benchmark}
\end{table}

To confirm the transferability of solvation models developed for plane wave electronic structure codes and different pseudopotentials to our GTO-based code, we computed single-point solvation energies for a benchmark set of 240 neutral molecules, 52 cations, and 60 anions.\cite{CCCBDB2022} This is also the dataset used for the calibration of previous solvation models.\cite{andreussi2012,gunceler2013,dupont2013dec} For all calculations, we used the PBE functional, and all molecules were placed in a 20 \AA\ cubic box with the GPW plane wave cutoff as 3000 eV. The results are shown in Table~\ref{tab:solv-benchmark}. The results suggest that these solvation energies do not have a strong dependence on basis set choice or pseudopotential.

The presence of electrolytes in our solvation model sets a natural reference electrostatic potential in the bulk electrolyte. Still, in most periodic electronic structure codes, the electrostatic potential is chosen such that the integral of the electrostatic potential is 0. Our implementation automatically shifts the reference potential, so one does not need to add a shifting quantity to the Fermi energy, such as EFERMI\_SHIFT in a VASPsol calculation.\cite{mathew2019} We verify our reference potential by plotting the grand canonical free energy near the PZC of an Ag(100) slab. We observe the expected parabolic shape to the energy curve as well as a free energy maximum at neutral charge.\cite{santos2004} The details of this are included in the Supporting Information.

\begin{figure}
    \centering
    \includegraphics[width=\linewidth]{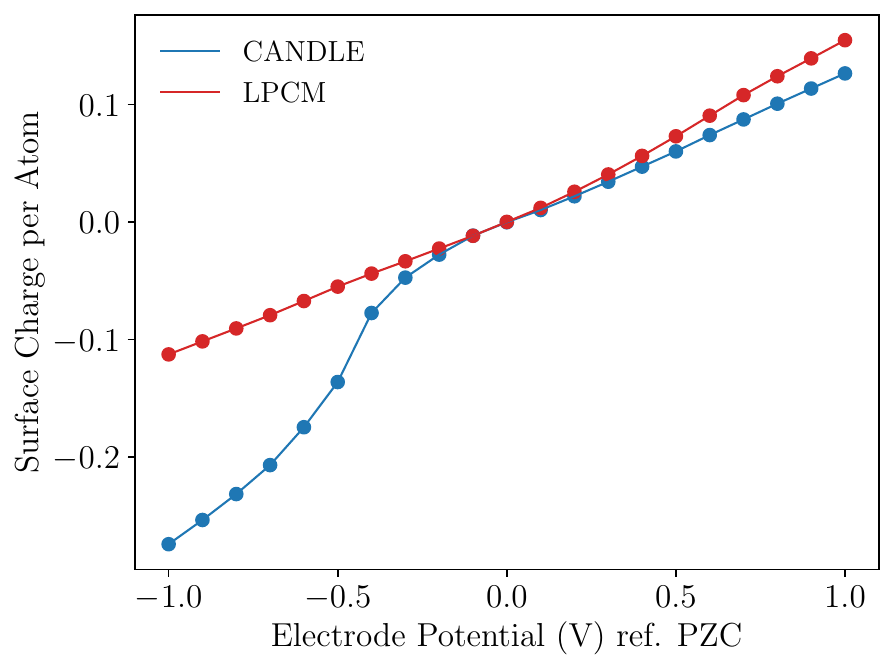}
    \caption{Surface charge as a function of electrode potential relative to PZC at 0.1 M ionic concentration with a computational unit cell height of 60.8 \AA~for LPCM and CANDLE.}
    \label{fig:ag100-charge}
\end{figure}

One potential weakness of implicit solvation models is the undercharging of electrodes as the electrode potential varies. Most such polarizable continuum models are fit to solvation energies of small molecules and ions. As a result, the capacitance of most metal electrodes is underestimated.\cite{sundararaman2017feb} Reparameterizations of preexisting solvation models have been presented to mitigate this issue, often at the cost of accuracy in energy evaluation.\cite{sundararaman2017feb} To combat this, we have implemented the CANDLE solvation model, which largely resolves this issue for negatively charged slabs by constructing a smaller solute cavity for the anionic regions of the solute. With our implementation of CANDLE, we were able to reproduce the results presented by Sundararaman and coworkers, where CANDLE accurately describes the charging behavior of negatively charged silver slabs.\cite{sundararaman2017feb} The results are presented in Fig.~\ref{fig:ag100-charge}. It should be noted that the capacitance is still underestimated for positively charged electrodes even with CANDLE.

\begin{figure}
    \centering
    \includegraphics[width=\linewidth]{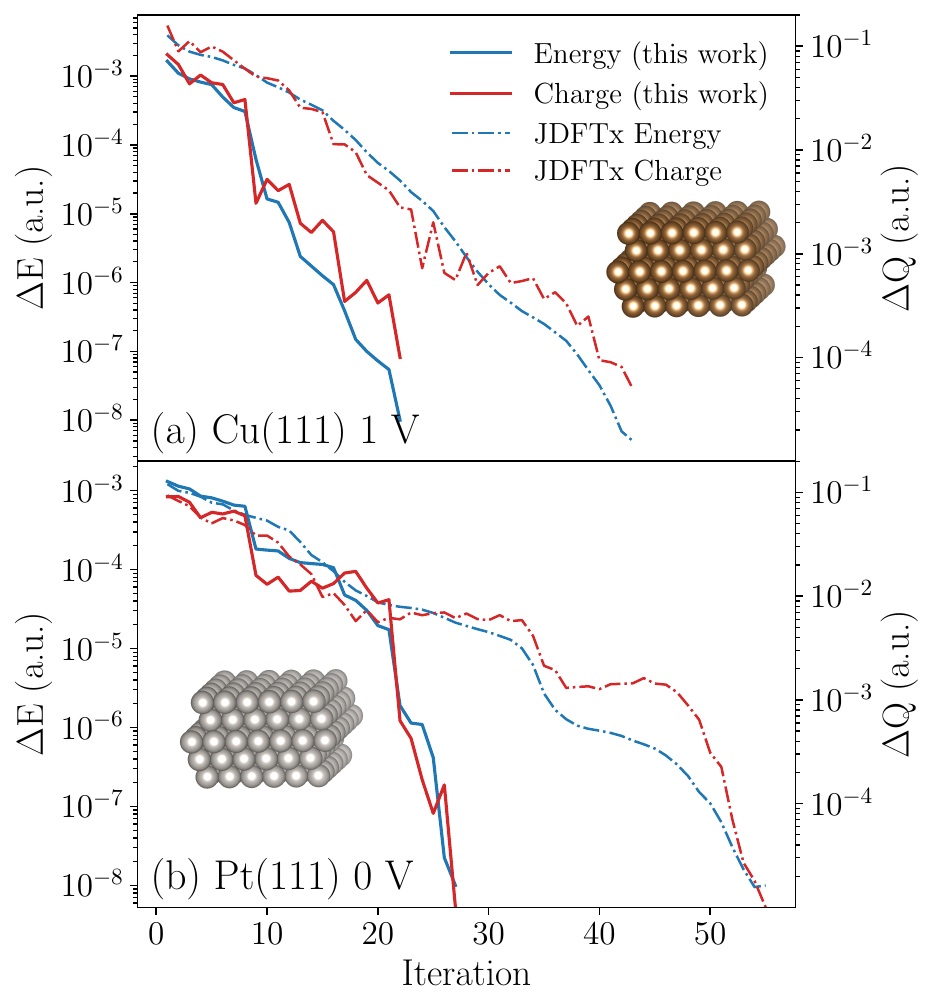}
    \caption{Energy ($\Delta E$) and charge ($\Delta Q$) convergence plot with CANDLE for two systems: (a) Cu(111) at 1 V
    SHE and (b) Pt(111) at 0 V SHE. The convergence is compared with results from JDFTx, which implements the auxiliary Hamiltonian method for minimization of the grand canonical free energy.\cite{sundararaman2017}} 
    \label{fig:convergence}
\end{figure}

\subsection{Grand canonical SCF convergence}
We compared the convergence behavior of our direct minimization approach to the auxiliary Hamiltonian method implemented in JDFTx, an efficient variational optimization approach for plane wave-based codes.\cite{sundararaman2017} 
In particular, we compared the SCF convergence against JDFTx for the two benchmark systems described in Sundararaman \textit{et. al.}: Cu(111) at 1 V and Pt(111) at 0 V relative to the computed standard hydrogen electrode (SHE).\cite{sundararaman2017,norskov2004} Consistent with their work, we use the CANDLE solvation model for this set of calculations. The results are shown in \cref{fig:convergence}. Our method demonstrates significantly better convergence over JDFTx, as we can converge energies and charge in about half the number of conjugate gradient iterations in both cases.
This provides evidence that our direct density matrix minimization approach is a robust strategy for grand canonical SCF calculations.

\subsection{Computational efficiency}
We also determine the computational overhead for including solvation in calculations. Ideally, one would like this overhead to be negligible compared to the rest of the computational components that are also common to canonical DFT. 
Because within each SCF iteration, we solve \cref{eq:pb} via the conjugate-gradient algorithm, we want to confirm that this iterative procedure does not significantly increase the runtime relative to the corresponding gas-phase calculation. 

\begin{figure}
    \centering
    \includegraphics[width=\linewidth]{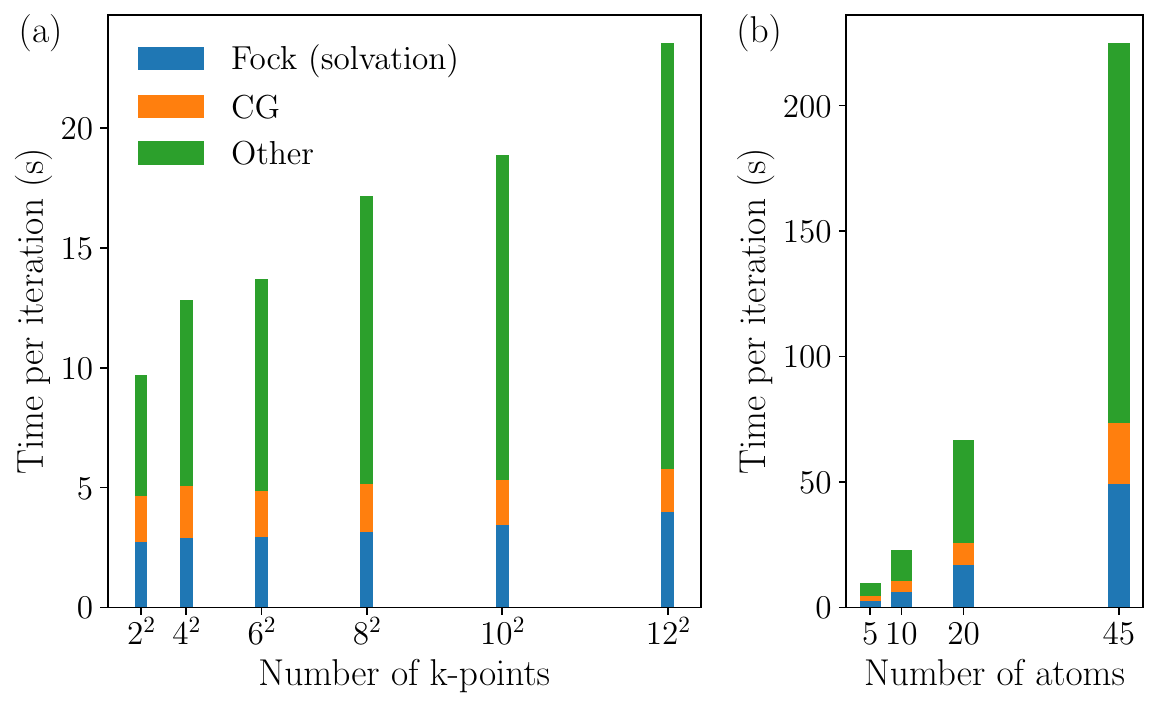}
    \caption{Timing breakdown as a function of (a) number of $\mathbf k$-points and (b) the number of atoms in the unit cell. The slab used was a Cu(100) slab with 15 \AA\ vacuum layer with the PBE functional. The sum of the blue (Fock) and orange (CG) represents the net solvation overhead.}
    \label{fig:solv-timings}
\end{figure}

We performed tests using a five-layer Cu(100) slab, and the results are shown in Fig.~\ref{fig:solv-timings}. 
We varied the number of $\mathbf k$-points in the calculation as well as the number of atoms in the unit cell by modulating the number of atoms per layer in the unit cell. 
Our implementation shows that the solvation introduces a less than 50\% increase in the cost per iteration, and the overhead scales minimally with the number of $\mathbf k$-points. This is promising for computations on metallic systems, where a large $\mathbf k$-mesh is typically required.

A direct comparison of overall timings between GTO- and plane wave-based codes can be challenging due to the differences in pseudopotentials, grid density, and basis set errors. Our solvation solver and related computational kernels are identical to those previously developed for plane waves.\cite{letchworth-weaver2012,mathew2019} Hence, the associated overhead for solvation is the same in our code and plane wave codes. The primary computational bottleneck in these calculations is the formation of the Fock matrix, which lies beyond the scope of this work, but we have recently reported progress.\cite{dinh2025sep} As a test, we also ran five-layer Cu(100) and Ag(100) slabs using our LPCM implementation and JDFTx on 32 threads on an AMD Ryzen 9 7950X CPU with the PBE functional. For the JDFTx calculations, we employed the recommended GBRV pseudopotentials with the associated energy cutoff of 20 $E_h$. In both cases, the calculations were initialized from a converged gas-phase electronic density. Our implementation required 79 s and 103 s to converge Cu(100) and Ag(100), respectively, while JDFTx required 77 s and 75 s. Although these timings are not directly comparable, they indicate that our GTO-based implementation achieves performance similar to existing plane wave implementations.

\subsection{Computational standard hydrogen electrode reference potential}
\begin{figure}
    \centering
    \includegraphics[width=\linewidth]{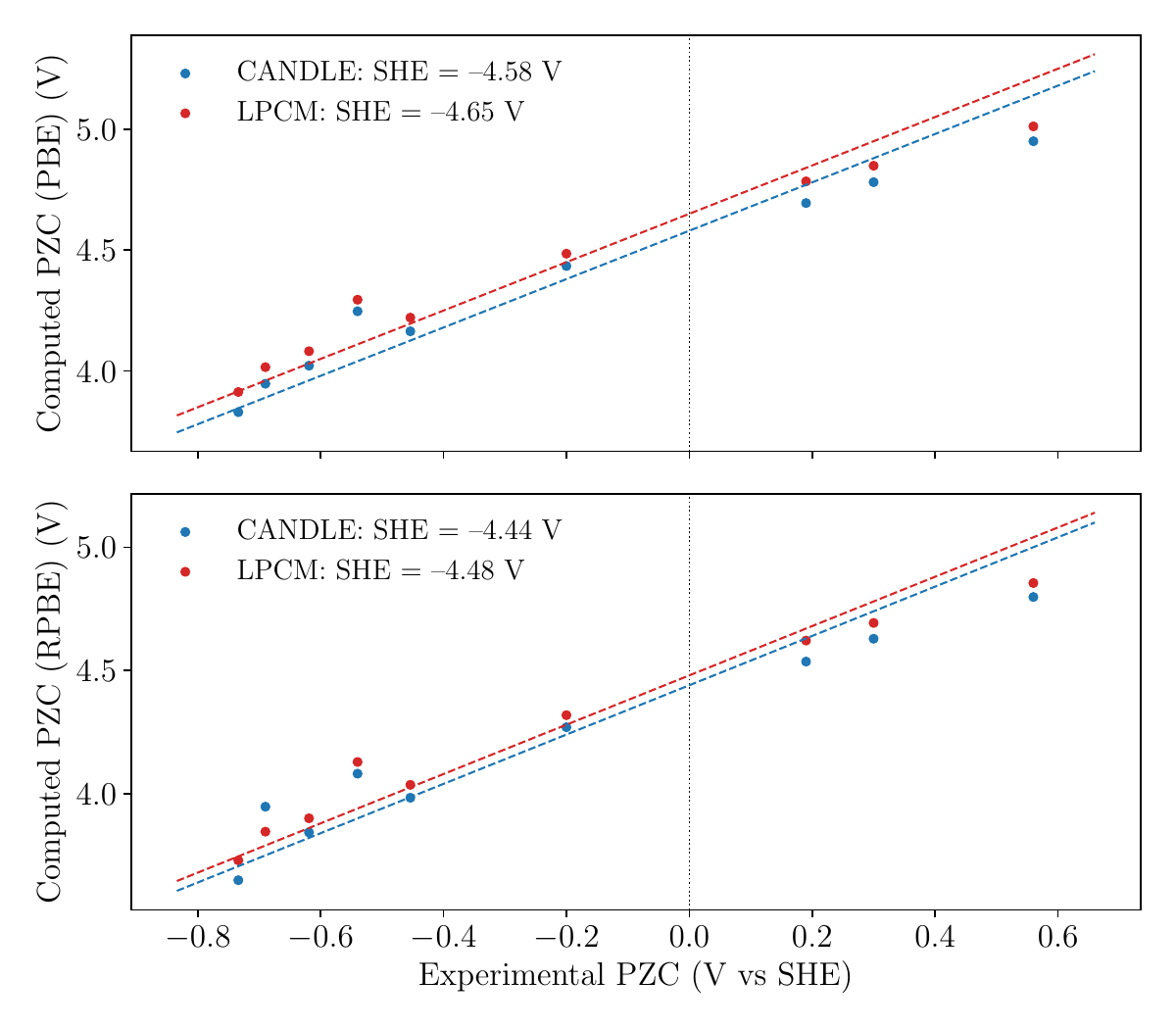}
    \caption{Potentials of zero charge predicted from LPCM and CANDLE along with the linear fit. Data points were obtained with (100), (110), (111) facets of Cu, Ag, and Au.}
    \label{fig:she}
\end{figure}

To perform constant potential calculations, an absolute electrode potential relative to vacuum must be established. In this work, we computed the absolute potential of the SHE relative to the vacuum using the method described by JDFTx.\cite{letchworth-weaver2012,gunceler2013} 
Experimental potentials of different facets of copper, silver, and gold are well benchmarked, so the computational potentials of zero charge (PZC) are fitted to these values to obtain the computational hydrogen electrode reference.\cite{TrasattiLust1999} These metals are chosen because they interact minimally with the electrolyte. For all calculations, 1.0 M of non-interacting electrolyte was assumed. The fitting obtained using our method is shown in Fig.~\ref{fig:she}. The computed SHE values using the PBE functional for the LPCM and CANDLE models are $-4.65$ V and $-4.58$ V, respectively. This value is in close agreement with values reported from other implementations of this method in VASP (LPCM: $-4.6$ V) and JDFTx (LPCM: $-4.68$ V, CANDLE: $-4.66$ V).\cite{gunceler2013,mathew2019}
We also fit SHE values for the RPBE functional that has been shown to give improved energetic results for absorption energies and other surface-related reactions.\cite{rpbe} We obtained computed SHE values of $-4.48$ V and $-4.44$ V for LPCM and CANDLE models respectively. 

\subsection{Practical study: silver corrosion}

We apply our method to study the corrosion of electrodes in aqueous interfacial chemistry. The surface transformations of electrified metal/solution interfaces are often critical to understanding electrochemical reaction pathways.\cite{gileadi2011} Recently, the thermodynamics and kinetics of anodic silver atom corrosion from an Ag(100) slab were extensively studied by Kang and coworkers.\cite{kang2024feb} Therein, the computations were performed using VASPsol, equivalent to the LPCM method we have implemented.\cite{mathew2019} 
To perform constant potential calculations in VASP, one must update the electron number in between SCF cycles until convergence, which can lead to additional computational and manual work.\cite{vaspsol++} 
Due to the body of computational and experimental work available for this system,
we believe that this is a well-established benchmark system to demonstrate the value of our method.

\begin{figure}
    \centering
    \includegraphics[width=\linewidth]{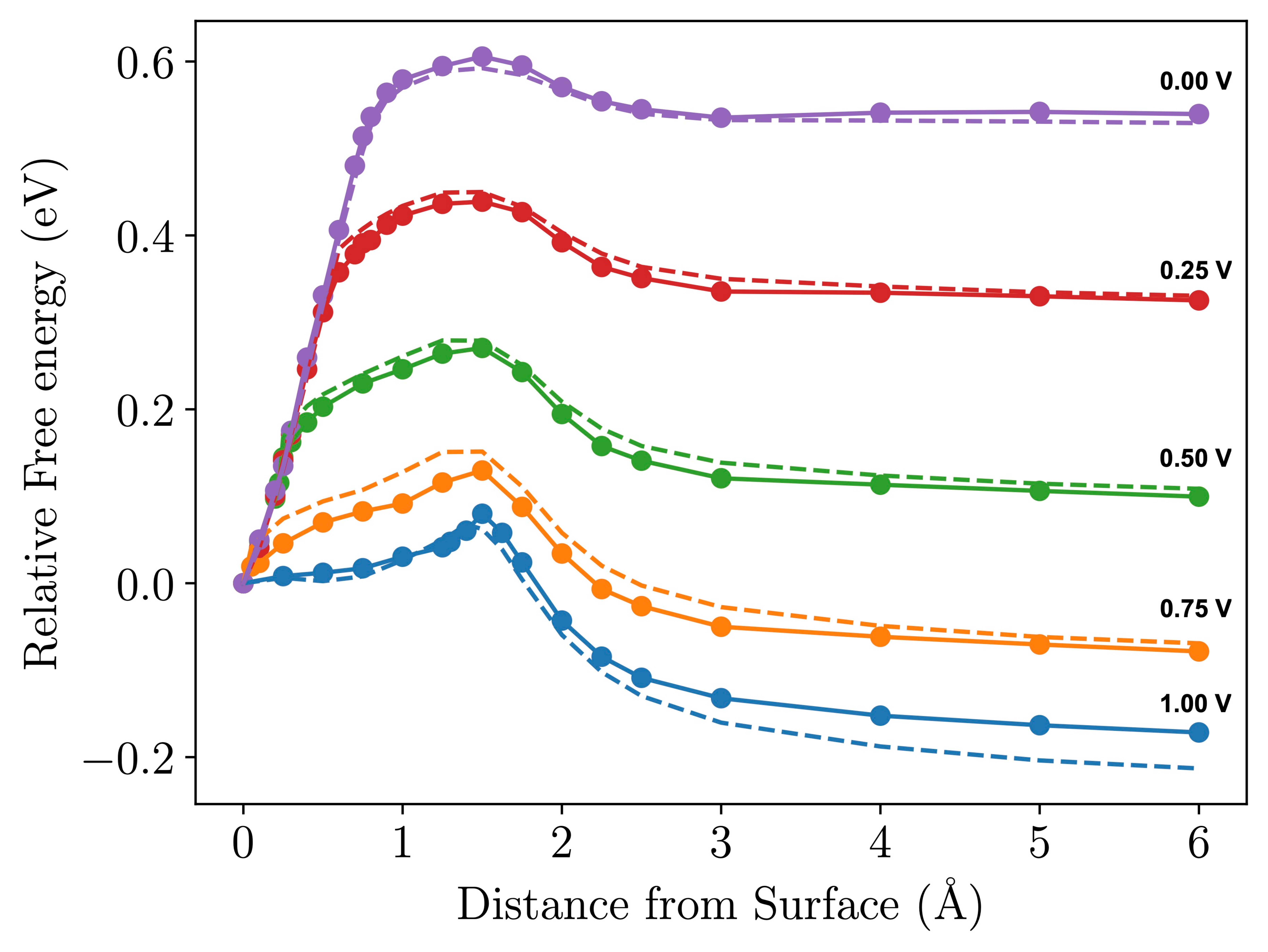}
    \caption{Potential energy curves for silver corrosion at different potentials relative to the computational SHE. The dashed lines shown are the results reported by Kang and coworkers.\cite{kang2024feb} All calculations were performed with the RPBE functional with a GPW plane-wave cutoff of 4800 eV except for at 1.00 V, where a GPW plane-wave cutoff of 6000 eV was used.}
    \label{fig:ag-corrosion}
\end{figure}

Here, we computed the thermodynamics of the corrosion of a single top-layer silver atom on an Ag(100) lattice at various electric biases as described in the work by Kang and coworkers, since undercoordinated sites are usually most susceptible to corrosion. 
Using our implementation of GCDFT coupled with LPCM, we were able to obtain the potential energy curves for the corrosion of an Ag(100) surface at five different electric biases (Fig.~\ref{fig:ag-corrosion}). 
For these calculations, we used the TZVP-MOLOPT-PBE-GTH basis set with the RPBE functional.\cite{rpbe} We used a unit cell containing 25 Ag atoms, consisting of a three-layer slab with an additional single-atom layer on top with 6$\times$6 $\mathbf k$-point sampling.
We also included two explicit water molecules providing the first coordination shell to the Ag atom coming off the surface.
The GPW plane-wave cutoff employed was 4800 eV with a unit-cell height of 60 \AA. Due to the slightly slower convergence of the plane wave cutoff at 1.00 V, we report data at 1.00 V with a GPW plane-wave cutoff of 6000 eV. The LPCM solvation model was used with cavitation turned off due to observed numerical instabilities.\cite{kang2024feb} 
The SHE reference used was calibrated to the experimental PZC of Ag(100) (0.609 V).\cite{TrasattiLust1999} We note that the reference PZC used in the original work was 0.603 V.

Our calculated equilibrium electrode potential was 0.70 V with an associated activation barrier of 0.20 eV. Both numbers are in good agreement with the numbers reported using VASPsol (0.68 V and 0.21 eV).\cite{kang2024feb} All further computational details are included in the Supporting Information. The computational tools used in this set of calculations can be applied to a wide range of other electrochemical systems.

\section{Conclusions}

This work presents the implementation of GCDFT with implicit solvation to facilitate computational electrochemistry calculations. We present a novel method to perform such calculations with GTOs. 
Our method leverages the compactness of the density matrix in a GTO basis, in contrast to a plane wave basis, and performs direct free energy minimization with respect to the density matrix. The implementation of this method is coupled with an implicit solvation model, which accounts for the presence of solvent and electrolytes, consistent with experimental electrochemistry setups. 
Our method demonstrates accelerated convergence relative to the current state-of-the-art methods for GCDFT. We use our implementation to confirm the results of a silver corrosion study from Kang and coworkers,~\cite{kang2024feb} which employs VASPsol,~\cite{mathew2019} one of the more widely used implementations for studying electrochemistry applications. 

Our method has the advantage of optimizing electron number within an SCF iteration, whereas constant potential calculations in VASPsol require updating electron number in between geometry optimization cycles. We also believe our implementation in a GTO-based code facilitates the usage of hybrid functionals~\cite{lee2021oct,lee2022dec,rettig2023sep} and correlated wavefunction methods~\cite{gruber2018may,carbone2024aug,wang2021oct,ye2024mar,zen2016nov,chen2025aug} for more accurate energies, directly benefiting from the entire body of work in quantum chemistry that accelerates these more sophisticated methods.
Overall, we believe the methods in this work will set the stage for enabling more realistic computational simulations of complex electrochemical environments with greater predictive power.

\begin{acknowledgments}
This work was funded by the startup funds given by Harvard University.
The work of A. Z. N. was also supported by the NSF GRFP Fellowship.
The work of A. R. was also supported by the Harvard Quantum Initiative prize postdoctoral fellowship. 
We thank Diptarka Hait for pointing out a typo in \cref{eq:discOmega} and Richard Kang for pointing out mislabels in \cref{fig:ag-corrosion}.
This work used computational resources from FASRC cluster supported by the Faculty of Arts and Sciences (FAS) Division of Science Research Computing Group at Harvard, and the Delta system at the National Center for Supercomputing Applications through allocation CHE250143 from the Advanced Cyberinfrastructure Coordination Ecosystem: Services \& Support (ACCESS) program, supported by National Science Foundation grants \#2138259, \#2138286, \#2138307, \#2137603, and \#2138296.
\end{acknowledgments}

\nocite{*}
\bibliography{refs}

\newpage
\appendix{}
\onecolumngrid
\renewcommand{\thefigure}{A\arabic{figure}}
\setcounter{figure}{0}
\begin{center}
    \textbf{Appendices}
\end{center}
\section{Details for the Direct Minimization Method}
\label{sec:gc-gradients}
This section will provide more details of our direct minimization method for grand canonical density functional theory. The grand canonical free energy we seek to minimize is 
\begin{equation}
    \Omega = E_\text{DFT} - \mu N - \frac{1}{\beta}S_\text{el},
\label{eq:si-grandpot}
\end{equation}
where $E_\text{DFT}$ is the DFT electronic energy, $\mu N$ is the electron number scaled by the fixed chemical potential $\mu$, $\beta$ is the inverse temperature, and $S_\text{el}$ is the electronic entropic energy.

$E_\text{DFT}$ consists of the one-body Hamiltonian energy (h), Coulomb energy (J), and DFT exchange-correlation energy (xc). We discretize the underlying operators with orthonormalized GTOs as defined in the manuscript: 
\begin{equation}
    \frac{\Omega}{N_k}=\frac{1}{N_k}\sum_{\mathbf k}\left(
    \Tr\!\big(\mathbf P^{\mathbf k}(\mathbf h^{\mathbf k}+\tfrac{1}{2}\mathbf J^{\mathbf k})\big)
    + E_{\mathrm{xc}}[\rho(\mathbf r),\dots]
    - \mu\,\Tr(\mathbf P^{\mathbf k})
    + \frac{1}{\beta} S_{\mathrm{el}}^{\mathbf k}
  \right).
  \label{eq:si-discOmega}
\end{equation}

As in Mermin's thermal Hartree-Fock theory, we want to minimize over all density matrices, Hermitian matrices with eigenvalues between 0 and 1.\cite{mermin1963} To turn this constrained minimization into an unconstrained minimization, we introduce our variational parameter $\{\mathbf{H}\}$, which we define such that $\mathbf{H^\k} = \mathbf{F^\k}$ at convergence. At convergence, we know that the electron population follows a Fermi-Dirac distribution, so we introduce $\bG^{\mathbf k} = \beta\left(\frac{1}{2}\left( {\H}^{\mathbf k} + {\H}^{\mathbf k H} \right) - \mu \textbf{I} \right) = \U^{\mathbf k}\bg^{\mathbf k}\U^{\mathbf k H}$ where $\U$ is the unitary matrix which diagonalizes $\bG$, and $\bg$ are the eigenvalues of $\bG$. This representation of $\bG$ also enforces the Hermiticity of $\H$. We can thus parameterize our density matrix as follows to satisfy the constraint that it must have eigenvalues between 0 and 1:
\begin{equation}
{\P}^{\mathbf k} = \frac{1}{\exp(\bG^{\mathbf k}) + 1} = \U^{\mathbf k}\frac{1}{\exp(\bg^{\mathbf k})+1}\U^{\mathbf k H} = \U^{\mathbf k}\boldrho^{\mathbf k}\U^{\mathbf k H}.
\label{eq:si-pfromh}
\end{equation}

Now we derive the gradients for direct minimization. To perform direct minimization, we need $\frac{\partial \Omega}{\partial \textbf{H}}$. We do this by differentiating with respect to both the real and imaginary components of $\textbf{H}$. To do this, we will differentiate with respect to the real and imaginary components of $\bG$. These gradients can subsequently be symmetrized to obtain $\frac{\partial \Omega}{\partial \H}$ by following the equations in the main text. We will first differentiate with respect to $E_\text{DFT}$.  Below, we use the fact that $\frac{\partial E_\text{DFT}}{\partial \mathbf{P}^\k} = \mathbf{F}^\k$.
\begin{align}
    \frac{\partial E_\text{DFT}}{\partial \Gamma^\k_{pq}} &= \frac{\partial E_\text{DFT}}{\partial P_{nm}^\k} \frac{\partial P_{nm}^\k}{\partial \Gamma^\k_{pq}} = F_{mn}^\k \frac{\partial (U_{mi}^\k\rho_{i}^\k U_{ni}^{\k*})}{\partial \Gamma_{pq}^\k}\\
    &= F^\k_{mn}\left( \frac{U_{mi}^\k}{\partial \Gamma_{pq}^\k}\rho_{i}^\k U_{ni}^{\k*} + U_{mi}^\k\rho_{i}^\k\frac{\partial U_{ni}^{\k *}}{\partial\Gamma_{pq}^\k} + U_{mi}^\k\frac{\partial \rho_i^\k}{\partial \Gamma_{pq}^\k}U_{ni}^{\k *} \right)
\end{align}
We use the fact that $\frac{\partial \rho}{\partial \gamma} = -\rho(1-\rho)$ and the form of the Hermitian eigenvector and eigenvalue derivatives described in the subsequent section to obtain the expression written in the main text:
\begin{align}
    \frac{\partial E_\text{DFT}}{\partial \text{Re}(\Gamma^\k_{pq})} &= {F}^\k_{mn}\left( U_{pl}^{\k *}U^\k_{qi}U_{nl}^\k G_{li}^\k \rho_i^\k U_{mi}^{\k *} + U_{ni}^\k \rho_i^\k U_{pl}^\k U_{qi}^{\k *} U_{ml}^{\k *} G_{li}^\k - U_{ni}^\k U_{mi}^{\k *}\rho_i^\k (1-\rho_i^\k)U_{pi}^{\k *} U_{qi}^{\k} \right)\\
    \frac{\partial E_\text{DFT}}{\partial \text{Im}(\Gamma^\k_{pq})} &= i{F}^\k_{mn}\left( U_{pl}^{\k *}U^\k_{qi}U_{nl}^\k G_{li}^\k \rho_i^\k U_{mi}^{\k *} - U_{ni}^\k \rho_i^\k U_{pl}^\k U_{qi}^{\k *} U_{ml}^{\k *} G_{li}^\k - U_{ni}^\k U_{mi}^{\k *}\rho_i^\k (1-\rho_i^\k)U_{pi}^{\k *} U_{qi}^{\k} \right).
\end{align}

The remaining two terms of Eq.~\ref{eq:si-grandpot} do not depend on the orbital rotations $\U$, so if we substitute in $\frac{\partial\rho}{\partial\gamma}$ and add back to $\frac{\partial E_\text{DFT}}{\partial \bG^\k}$, we get the equations in the main text:
\begin{gather}
    \begin{split}
        \frac{\partial \Omega}{\partial \text{Re}(\Gamma^\k_{pq})} =\quad &{F}^\k_{mn}\left( U_{pl}^{\k *}U^\k_{qi}U_{nl}^\k G_{li}^\k \rho_i^\k U_{mi}^{\k *} + U_{ni}^\k \rho_i^\k U_{pl}^\k U_{qi}^{\k *} U_{ml}^{\k *} G_{li}^\k - U_{ni}^\k U_{mi}^{\k *}\rho_i^\k (1-\rho_i^\k)U_{pi}^{\k *} U_{qi}^{\k} \right) \\
        &+ \mu (U_{pi}^{\k *} \rho_i^\k (1-\rho_i^\k)U_{qi}^{\k}) + \frac{1}{\beta}\gamma_i^\k \rho_i^\k(1-\rho_i^\k)U_{pi}^{\k *} U_{qi}^{\k}
    \end{split}\\
    \begin{split}
        \frac{\partial \Omega}{\partial \text{Im}(\Gamma^\k_{pq})} = \quad  &i{F}^\k_{mn}\left( U_{pl}^{\k *}U^\k_{qi}U_{nl}^\k G_{li}^\k \rho_i^\k U_{mi}^{\k *} - U_{ni}^\k \rho_i^\k U_{pl}^\k U_{qi}^{\k *} U_{ml}^{\k *} G_{li}^\k - U_{ni}^\k U_{mi}^{\k *}\rho_i^\k (1-\rho_i^\k)U_{pi}^{\k *} U_{qi}^{\k} \right) \\
        &+ i\mu (U_{pi}^{\k *} \rho_i^\k (1-\rho_i^\k)U_{qi}^{\k}) + i\frac{1}{\beta}\gamma_i^\k \rho_i^\k(1-\rho_i^\k)U_{pi}^{\k *} U_{qi}^{\k}
    \end{split}
\end{gather}

\subsection{Eigenvalue and Eigenvector Derivatives of Hermitian Matrices}
\label{subsec:eval-deriv}
We first derive a generalized form for eigenvalue and eigenvector derivatives for eigenvalues $\text{\boldmath$\lambda$}$ and eigenvectors \textbf{U} of some Hermitian matrix \textbf{H}. We denote a singular eigenvector by $u_i$, and the derivative with respect to some parameters as $\dot{u}_i$, and we define eigenvalues analogously using $\lambda_i$. We first derive the general form of the complex eigenvalue and eigenvector derivative assuming only that the eigenvectors are orthonormal; here we define $\C:= \U^H \dot{\U}$:
\begin{align*}
    \H\U &= \U\bL \\
    \dot{\H}\U - \U\dot{\bL} &= -\H\dot{\U} + \dot{\U}\bL \\
    \U^H\dot{\H}\U - \dot{\bL} &= \bL\C + \C\bL \\
    u_l^H\dot{\H}u_i - \delta_{li}\dot{\lambda}_i &= (\lambda_i - \lambda_l)c_{li}
\end{align*}
From the above derivation, we obtain a few facts:
\begin{align}
    \dot{\lambda}_i &= u_i^H \dot{\H} u_i\\
    \dot{\U} &= \U\C \label{eq:si-UC}\\
    c_{li} &= \frac{u_l^H\dot{\H}u_i}{\lambda_i - \lambda_l}\quad \text{ if $l\neq i$} \\
    \text{Re}(c_{ii}) &= 0
\end{align}

When diagonalizing a Hermitian matrix, the eigenvectors are typically only subject to the orthonormality constraint, so they are specified up to some arbitrary phase $\exp(i\theta)$. This nondeterminacy shows up in the derivatives derived above because $\text{Im}(c_{ii})$ is ill-defined. However, because the energy is gauge-invariant, all instances of $c_{ii}$ will necessarily cancel out in the energy derivative, so in our implementation, we set $c_{ii}=0$.

\section{Derivation of Kohn-Sham Potential due to Solvation}
\label{sec:vsolv-derivation}
Here, we describe the potential correction that needs to be added to the Kohn-Sham mean-field potential. We rewrite the general energy expression for the solvation energy in terms of linear dielectric and ionic response. The solvation energy expression can be written as \begin{equation}
    E_\text{solv} = E_{J,\text{solv}}[\rho,\phi] + E_\text{diel}[\rho,\phi] + E_\text{ion}[\rho,\phi] + E_\text{cd}[\rho] 
\end{equation} The correction to the Kohn-Sham mean-field potential is given by \begin{equation}
    V_\text{solv}(\textbf{r}) = \frac{\delta E_\text{solv}}{\delta \rho(\textbf{r})}\label{eq:si-app-vsolv}
\end{equation}
For the rest of this section, we will assume \textbf{r} dependence on all quantities unless otherwise specified.

\noindent We first derive the form of the Coulomb correction energy presented in the manuscript. Given the electrostatic potential $\phi$, we have \begin{equation}
    E_{J,\text{solv}} = \int d^3\textbf{r} \rho_\text{sol}\phi -\frac{1}{8\pi}(\nabla\phi \cdot \nabla\phi) - \frac{1}{2}\rho_\text{sol}\phi_\text{sol}.
\end{equation}
In this form, it is easy to see that $\frac{\delta E_{J,\text{solv}}}{\delta\rho} = \phi_\text{solv}$.\newline
Using the Poisson's equation $\nabla^2\phi_\text{sol} = -4\pi\rho_\text{sol} + 4\pi \frac{q_\text{sol}}{V}$, $\int d^3\textbf{r}\; \phi_\text{sol} = 0$, and integrating by parts, we get that 
\begin{align}
    E_{J,\text{solv}} &= \int d^3\textbf{r}\; \frac{1}{2}\rho_\text{sol}\phi_\text{sol} + \rho_\text{sol}\phi_\text{solv} + \frac{1}{8\pi}\phi\cdot\nabla^2\phi \\
    &= \int d^3\textbf{r}\; \frac{1}{2}\rho_\text{sol}\phi_\text{sol} + \rho_\text{sol}\phi_\text{solv} + \frac{1}{2}\left(\frac{q_\text{sol}}{V} - \rho_\text{sol}\right)\phi + \frac{1}{8\pi}\phi_\text{solv}\nabla^2\phi_\text{sol} + \frac{1}{8\pi}\phi_\text{solv}\nabla^2\phi_\text{solv} \\
    &= \int d^3\textbf{r}\; \frac{1}{2}\rho_\text{sol}\phi_\text{solv} + \frac{1}{2}\frac{q_\text{sol}}{V}\phi_\text{solv} + \frac{1}{2}\left(\frac{q_\text{sol}}{V} - \rho_\text{sol} \right)\phi_\text{solv} + \frac{1}{8\pi}\phi_\text{solv}\nabla^2\phi_\text{solv} \\
    &= \frac{q_\text{sol}}{V}\int d^3\textbf{r}\; \phi_\text{solv} + \frac{1}{8\pi}\int d^3\textbf{r}\; \phi_\text{solv}\nabla^2\phi_\text{solv}
\end{align}
as shown in the manuscript.

The dielectric and ionic free energies for both LPCM and CANDLE take the same form: \begin{equation}
    E = \int d^3\textbf{r}\; s_\epsilon[\rho,\nabla\rho] \cdot \lambda_\text{lq}[\phi]
\end{equation}
\begin{equation}
    \lambda_\text{lq}(\textbf{r}) = -\frac{(\epsilon_b-1)}{8\pi}|\nabla\phi(\textbf{r})|^2 -\frac{\epsilon_b\kappa^2}{8\pi}\phi(\textbf{r})^2
\end{equation}
Therefore, the corresponding component of $V_\text{solv}$, $V_\text{lq}$, can be written as $\frac{\delta s_\epsilon}{\delta\rho}\lambda_\text{lq}$. We derive the specific forms of $V_\text{lq}$ for the two solvation models as well as the general expression to obtain the nonelectrostatic potential in the following two sections.

\subsection{LPCM Solvation Model}
\noindent The cavity function in LPCM is given by 
\begin{equation}
    s_{\epsilon,\text{LPCM}}(\textbf{r}) = \frac{1}{2}\text{erfc}\left( \frac{\log(\rho(\textbf{r})/\rho_\text{cut})}{\sigma\sqrt{2}} \right)
\end{equation}
Because $s_\epsilon$ only depends on $\rho(\textbf{r})$, we have
\begin{equation}
    \frac{\delta s_\epsilon}{\delta\rho} = \frac{\partial s_\epsilon}{\partial\rho} = -\frac{1}{\sqrt{2\pi}\sigma\rho(\textbf{r})}\exp\left( - \frac{\log^2(\rho(\textbf{r})/\rho_\text{cut})}{2\sigma^2} \right)
\end{equation}

\noindent Now we derive the cavitation potential. Using the fact that the shape function $s$ is monotonically decreasing in $\rho$, we can write the cavitation energy as \begin{equation}
    E_\text{cav, LPCM}[\rho] = \tau\int d^3\textbf{r}\; |\nabla s(\textbf{r})| = -\tau\int d^3\textbf{r}\; \frac{\partial s}{\partial\rho}|\nabla \rho(\textbf{r})|.
\end{equation}
Now we derive $V_\text{cav, LPCM}$. We denote $\partial\rho_{i}$ as the $i$-th component of the spatial gradient of $\rho$, with successive subscripts representing higher order spatial gradients.\cite{andreussi2012}
\begin{align}
    V_\text{cav, LPCM} &= \tau\frac{\partial s}{\partial\rho}\left(\frac{\partial}{\partial\nabla\rho}\cdot|\nabla\rho|\right) \\
    &= \tau\frac{\partial s}{\partial\rho}\left( \frac{\partial\rho_{xx}+\partial\rho_{yy}+\partial\rho_{zz}}{|\nabla\rho|} - \frac{\sum_i\partial\rho_i(\partial\rho_x\partial\rho_{xi} + \partial\rho_y\partial\rho_{yi} + \partial\rho_z\partial\rho_{zi}) }{|\nabla\rho|^3} \right)
\end{align}

\subsection{CANDLE Solvation Model}
We write the CANDLE cavity functions here again, which are obtained from the ``solvent-weighted'' electronic density $\overline{\rho} = w_\text{lq}*\rho$, where $w_\text{lq}$ is a Gaussian which has a solvent-dependent width:\cite{candle} 
\begin{equation}
    s_\text{CANDLE}(\textbf{r}) = \frac{1}{2}\text{erfc}\left( \log\frac{Z_\text{val}\overline{\rho}(\textbf{r})}{0.00142} - \text{sign}(p_\text{cav})f_\text{sat}\left( |p_\text{cav}|\hat{e}_{\nabla\overline{\rho}} \cdot \nabla\overline{\phi}_\text{sol} \right) \right)
\end{equation}
\begin{equation}
    f_\text{sat}(x) =
\begin{cases}
0, & x < 0, \\
3\tanh\left(x^2\right), & x \ge 0
\end{cases}
\end{equation}
\begin{equation}
    s_{\epsilon,\text{CANDLE}}(\textbf{r}) = \frac{\delta(|\textbf{r}| - R_{\epsilon})}{4\pi R_{\epsilon}^2} * s_\text{CANDLE}(\textbf{r}) \equiv \delta_{R_\epsilon} * s_\text{CANDLE}(\textbf{r})
\end{equation}

\noindent We note that $s_\text{CANDLE}$ has three terms dependent on $\rho$: $\overline{\rho}$, $\hat{e}_{\nabla\overline{\rho}}$, and $\nabla\overline{\phi}_\text{sol}$. 
We have \begin{gather}
    \frac{\partial s}{\partial \overline{\rho}} = -\frac{1}{\overline{\rho}\sqrt{\pi}}\exp\left(- \left( \log\frac{Z_\text{val}\overline{\rho}(\textbf{r})}{0.00142} - \text{sign}(p_\text{cav})f_\text{sat}\left( |p_\text{cav}|\hat{e}_{\nabla\overline{\rho}} \cdot \nabla\overline{\phi}_\text{sol} \right) \right)^2 \right) \equiv -\frac{1}{\overline{\rho}\sqrt{\pi}}e^{-X^2}\\
    \frac{\partial s}{\partial \nabla\overline{\rho}} = \frac{1}{|\nabla\overline{\rho}|\sqrt{\pi}}\exp\left(-X^2\right)p_\text{cav}f_\text{sat}'\left( |p_\text{cav}|\hat{e}_{\nabla\overline{\rho}} \cdot \nabla\overline{\phi}_\text{sol} \right)\cdot\left( \nabla\overline{\phi}_\text{sol} - \hat{e}_{\nabla\overline{\rho}}\left( \hat{e}_{\nabla\overline{\rho}}\cdot \nabla\overline{\phi}_\text{sol}\right) \right) \\
    \frac{\partial s}{\partial\nabla\overline{\phi}_\text{sol}} = \frac{1}{\sqrt{\pi}}\exp\left( -X^2 \right)p_\text{cav}f_\text{sat}'\left( |p_\text{cav}|\hat{e}_{\nabla\overline{\rho}} \cdot \nabla\overline{\phi}_\text{sol} \right)\cdot \hat{e}_{\nabla\overline{\rho}}
\end{gather}

The functional derivative of $\overline\rho$ with respect to $\rho$ is the kernel which applies the convolution with $w_\text{lq}$. Since $\nabla^2\overline{\phi}_\text{sol} = -4\pi\overline{\rho}_\text{sol}$, the functional derivative of $\overline{\phi}_\text{sol}$ with respect to $\rho$ is the kernel which applies the inverse Coulomb operator followed by the convolution with $w_\text{lq}$. We call this operator $\hat{K}_w*$. Because we are working in periodic boundary conditions, this has a convenient representation in reciprocal space (\textbf{G}-space): 
\begin{equation}
    \hat{K}_w*= \frac{\delta\overline{\phi}_\text{sol}}{\delta \rho}(\textbf{G}) = \frac{4\pi}{|\textbf{G}|^2}w_\text{lq} (\textbf{G}).
\end{equation}

We can now derive the dielectric and ionic contributions to $V_\text{solv}$. Here, we denote $\delta_{R_\epsilon} * \lambda_\text{lq}$ as $\tilde{\lambda}_\text{lq}$:
\begin{equation}
    E_\text{lq} = \int d^3\textbf{r}\; s_\epsilon(\textbf{r})\lambda_\text{lq}(\textbf{r}) = \int d^3\textbf{r}\; (\delta_{R_\epsilon} *s)(\textbf{r})\cdot \lambda_\text{lq}(\textbf{r})
\end{equation}
\begin{align}
    V_\text{lq}(\textbf{r}) &= \frac{\delta (E_\text{diel} + E_\text{ion})}{\delta \rho(\textbf{r})} \\
    &= \int d^3 \textbf{r}' \int d^3 \textbf{r}'' \frac{\delta(|\textbf{r}' - \textbf{r}''|- R_\epsilon)}{4\pi R_\epsilon^2}\lambda_\text{lq}(\textbf{r}')\frac{\delta s(\textbf{r}'')}{\delta\rho(\textbf{r})} \\
    &= \int d^3 \textbf{r}''\; \tilde{\lambda}_\text{lq} (\textbf{r}'') \frac{\delta s(\textbf{r}'')}{\delta\rho(\textbf{r})} \\
    &= w_\text{lq} * \left( \tilde{\lambda}_\text{lq} \cdot \frac{\partial s}{\partial \overline{\rho}} - \nabla\cdot\left(\tilde{\lambda}_\text{lq} \cdot \frac{\partial s}{\partial \nabla\overline{\rho}} \right) \right) - \hat{K}_w * \nabla\cdot\left( \tilde{\lambda}_\text{lq} \cdot \frac{\partial s}{\partial \nabla\overline{\phi}_\text{sol}} \right)\label{eq:si-candle_vsolvel}
\end{align}
All convolutions can be evaluated efficiently in reciprocal space.

The detailed expressions for cavitation and dispersion have previously been extensively detailed.\cite{sundararaman2014oct,candle} The cavitation energy can be generalized to the following form, where $\overline{s}$ is the solvent-expanded shape function $\delta_{\sigma_\text{vdW}}* s$ where $\sigma_\text{vdW}$ is the solvent vdW diameter.
\begin{equation}
    E_\text{cav} = \int d^3\textbf{r}\; f(\overline{s}).
\end{equation}
Here, $f$ is some polynomial function which depends only on $\overline{s}$ and some solvent-dependent constants.

\noindent If we write $\delta_{\sigma_\text{vdW}}*f'(\overline{s})$ as $\tilde{f'}(\overline{s})$, similarly to equation~\ref{eq:si-candle_vsolvel}, we can write \begin{equation}
    V_\text{cav} = \frac{\delta E_\text{cav}}{\delta \rho(\textbf{r})} = w_\text{lq} * \left( \tilde{f'}(\overline{s}) \cdot \frac{\partial s}{\partial \overline{\rho}} - \nabla\cdot\left(\tilde{f'}(\overline{s}) \cdot \frac{\partial s}{\partial \nabla\overline{\rho}} \right) \right) - \hat{K}_w * \nabla\cdot\left( \tilde{f'}(\overline{s}) \cdot \frac{\partial s}{\partial \nabla\overline{\phi}_\text{sol}} \right)
\end{equation} 

\noindent The dispersion energy is computed from the solvent-weighted cavity function $w_\text{lq}*s$. It takes the form \begin{equation}
    E_\text{disp} = \int d^3\textbf{r}\; (w_\text{lq}*s(\textbf{r})) \cdot \lambda_\text{disp}(\textbf{r})
\end{equation}

\noindent Therefore, writing $w_\text{lq}*\lambda_\text{disp}$ as $\tilde{\lambda}_\text{disp}$, we have \begin{equation}
    V_\text{disp} = w_\text{lq} * \left( \tilde{\lambda}_\text{disp} \cdot \frac{\partial s}{\partial \overline{\rho}} - \nabla\cdot\left(\tilde{\lambda}_\text{disp} \cdot \frac{\partial s}{\partial \nabla\overline{\rho}} \right) \right) - \hat{K}_w * \nabla\cdot\left( \tilde{\lambda}_\text{disp} \cdot \frac{\partial s}{\partial \nabla\overline{\phi}_\text{sol}} \right)
\end{equation}

\section{Verification of Absolute Vacuum Potential}
\label{sec:vacuum-potential}
\begin{figure}
    \centering
    \includegraphics[width=0.5\linewidth]{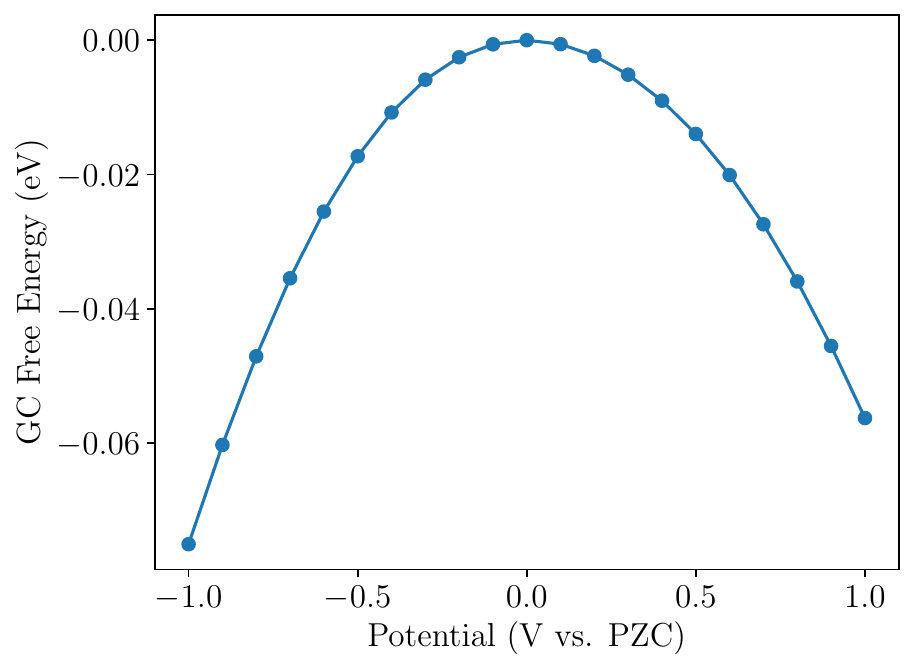}
    \caption{Plot of $E - \mu q_\text{sol}$ for an Ag(100) slab as a function of target chemical potential. We note the maximum in the grand canonical free energy at the PZC and the parabolic curve.}
    \label{fig:Eref}
\end{figure}

In most periodic electronic structure codes, the electrostatic potential is typically chosen such that the average of the electrostatic potential is 0. However, the presence of electrolytes in our solvation model sets a natural reference electrostatic potential in the bulk of the electrolyte. In our implementation, we automatically shift this reference potential, so one does not need to add an EFERMI\_SHIFT as in a VASPsol calculation.\cite{mathew2019} We verify our reference potential by plotting the quantity $E - \mu q_\text{sol}$, where $\mu$ is the target chemical potential, and $q_\text{sol}$ is the solute charge per unit cell at the potential $\mu$. If our reference potential is correct, we should observe a parabolic curve with a maximum at $q_\text{sol} = 0$.\cite{santos2004} Shown in Fig~\ref{fig:Eref} is a sample curve for the Ag(100) surface used in the main text. 

\section{Fitting Potentials of Zero Charge (PZC)}
\noindent The raw data for the fitting of TZCs for the PBC and RPBE functionals are in Table~\ref{tab:pzc-raw}. All numbers are in eV.
\begin{table}
\caption{Calculated and experimental potential of zero charge (PZC) values
    for Ag, Au, and Cu facets (in V vs. SHE).}
    \label{tab:pzc-raw}
    \begin{tabular}{lccccr}
    \hline
    \multirow{2}{*}{Species} & \multicolumn{2}{c}{PBE} & \multicolumn{2}{c}{RPBE} & \multirow{2}{*}{Exp. PZC (vs. SHE)} \\
    \cline{2-3} \cline{4-5}
    & LPCM & CANDLE & LPCM & CANDLE & \\
    \hline
    Ag(100) & --4.08175 & --4.02180 & --3.90037 & --3.84315 & --0.619 \\
    Ag(110) & --3.91316 & --3.82985 & --3.72973 & --3.64922 & --0.734 \\
    Ag(111) & --4.22066 & --4.16408 & --4.03619 & --3.98370 & --0.454 \\
    Au(100) & --4.84862 & --4.78084 & --4.69277 & --4.62868 &   0.300 \\
    Au(110) & --4.78412 & --4.69434 & --4.62134 & --4.53597 &   0.190 \\
    Au(111) & --5.01140 & --4.95005 & --4.85460 & --4.79781 &   0.560 \\
    Cu(100) & --4.29484 & --4.24692 & --4.12889 & --4.08129 & --0.540 \\
    Cu(110) & --4.01590 & --3.94733 & --3.84623 & --3.94733 & --0.690 \\
    Cu(111) & --4.48515 & --4.43391 & --4.31866 & --4.26965 & --0.200 \\
    \hline
    \end{tabular}
\end{table}

\section{Computational Details of the Ag Corrosion Calculations}
\label{sec:appendix-ag}
\begin{figure}
    \centering
    \includegraphics[width=0.5\linewidth]{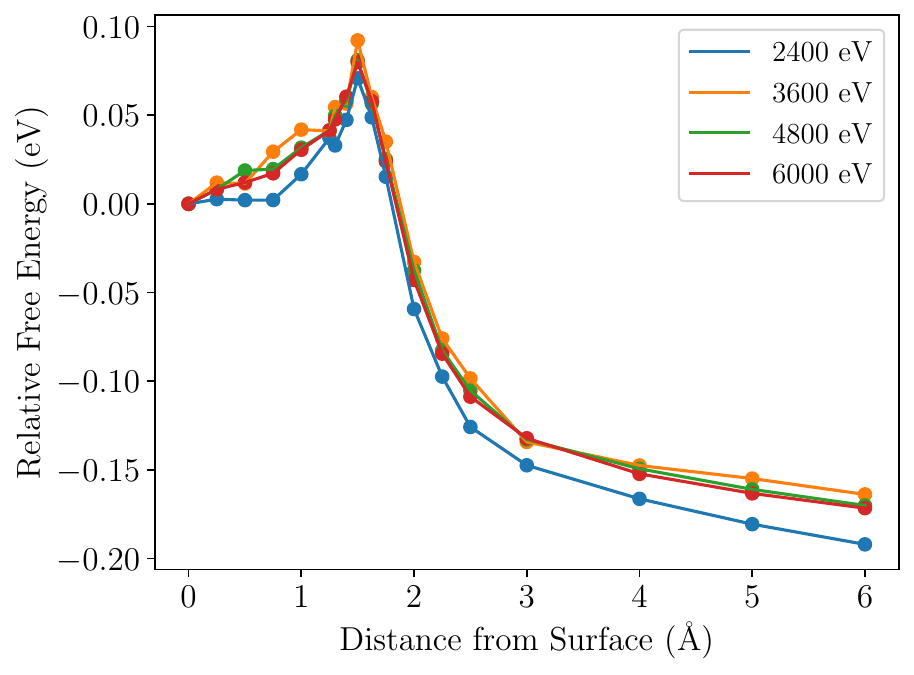}
    \caption{GPW plane wave cutoff determination for the Ag corrosion system at 1.00 V. A unit cell height of 60 \AA\, was used for these calculations.}
    \label{fig:ag25-kecut}
\end{figure}

The geometries for all of the calculations are from prior work by Kang and coworkers.\cite{kang2024feb} We first converged the GPW plane wave cutoff for our calculations. Shown in Fig.~\ref{fig:ag25-kecut} are the results for 1.00 V, which required the largest cutoff to converge fully. We note that the relative energies change minimally when increasing the GPW plane wave density fitting cutoff from 4800 eV to 6000 eV for 1.00 V. In the main text, we use 4800 eV as the GPW plane wave cutoff for 0.00 V, 0.25 V, 0.50 V, and 0.75 V. To ensure full plane wave cutoff convergence for 1.00 V, 6000 eV was used.

\begin{figure}
    \centering
    \includegraphics[width=0.5\linewidth]{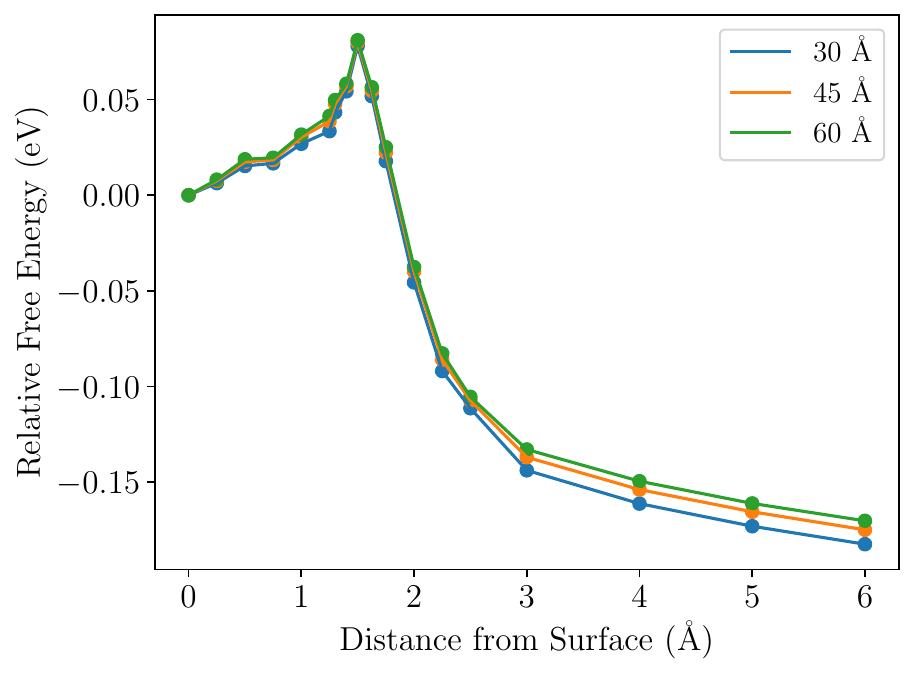}
    \caption{Potential energy curves for the Ag corrosion system at 1.00 V. All of these calculations were performed with a GPW plane wave density fitting cutoff of 4800 eV.}
    \label{fig:ag-box-height}
\end{figure}

Next, we obtained potential energy curves at multiple unit cell heights to ensure that the finite-size error was minimized. We sampled unit cell heights of 30 \AA, 45 \AA, and 60 \AA. The results are shown in Fig.~\ref{fig:ag-box-height}. We note that the barrier heights change minimally, but the change in cell height affects the free energy change of the reaction slightly. We elected to use a box height of 60 \AA\, for all calculations. The box size was chosen such that the change in the relative free energies was below $10^{-5}$ Hartree per atom in the unit cell.

Finally, the equilibrium potential was obtained via a least squares regression fit of the five reaction energies probed. Then, a linear fit of the computed activation barriers was used to determine the activation barrier at the predicted equilibrium potential. The details are available in the attached GitHub repository, which includes all the raw data.

\section{Raw Data}
All the Q-Chem inputs, raw data, and plotting scripts for this work can be found at the following GitHub repository: https://github.com/JoonhoLee-Group/gto-based-gcdft-data.

\end{document}